\documentstyle[12pt,twoside]{article}
{\catcode `\@=11 \global\let\AddToReset=\@addtoreset}
\AddToReset{equation}{section}
  
\newcommand{\sgn}{{\rm sgn}}
\def\greaterthansquiggle{\raise.3ex\hbox{$>$\kern-.75em\lower1ex\hbox{$\sim$}}}
\def\lessthansquiggle{\raise.3ex\hbox{$<$\kern-.75em\lower1ex\hbox{$\sim$}}}
\newcommand{\beq}{\begin{equation}}
\newcommand{\eeq}{\end{equation}}
\newcommand{\beqa}{\begin{eqnarray}}
\newcommand{\eeqa}{\end{eqnarray}}
\newcommand{\beqan}{\begin{eqnarray*}}
\newcommand{\eeqan}{\end{eqnarray*}}
\newcommand{\ba}{\begin{array}}
\newcommand{\ea}{\end{array}}
\newcommand{\no}{\nonumber}

\newcommand{\ol}{\overline}
\newcommand{\ra}{\rightarrow}

\newcommand{\dar}{\downarrow}
\newcommand{\ve}{\varepsilon}
\newcommand{\vp}{\varphi}

\newcommand{\wt}{\widetilde}

\newcommand{\A}{{\cal A}}
\newcommand{\B}{{\cal B}}
\newcommand{\C}{{\cal C}}
\newcommand{\D}{{\cal D}}

\newcommand{\Ha}{{\cal H}}

\newcommand{\st}{\stackrel}

\newcommand{\dsum}{\displaystyle \sum}
\newcommand{\dprod}{\displaystyle \prod}

\def\nz{\ifmmode {I\hskip -3pt N} \else {\hbox {$I\hskip -3pt N$}}\fi}
\def\zz{\ifmmode {Z\hskip -4.8pt Z} \else
       {\hbox {$Z\hskip -4.8pt Z$}}\fi}
\def\qz{\ifmmode {Q\hskip -5.0pt\vrule height6.0pt depth 0pt
       \hskip 6pt} \else {\hbox
       {$Q\hskip -5.0pt\vrule height6.0pt depth 0pt\hskip 6pt$}}\fi}
\def\rz{\ifmmode {I\hskip -3pt R} \else {\hbox {$I\hskip -3pt R$}}\fi}
\def\cz{\ifmmode {C\hskip -4.8pt\vrule height5.8pt\hskip 6.3pt} \else
       {\hbox {$C\hskip -4.8pt\vrule height5.8pt\hskip 6.3pt$}}\fi}

\def\lint{\int\limits}
\normalsize
\begin{document}
\bibliographystyle{plain}
\begin{titlepage}
\begin{flushright}
UWThPh-1998-37\\
ESI--587--1998\\
July, 1998
\end{flushright}

\vspace*{0.9cm}
\begin{center}
{\Large \bf  
The Thirring Model 40 Years Later$^\star$}\\[25pt]
N. Ilieva$^{\ast,\sharp}$ and W. Thirring  \\ [10pt]
Institut f\"ur Theoretische Physik \\ Universit\"at Wien\\
\smallskip
and \\
\smallskip
Erwin Schr\"odinger International Institute\\
for Mathematical Physics\\
\vspace{0.8cm}
{\bf Abstract} \\ [3pt]
\end{center}
Solutions to the Thirring model are constructed in the framework of
algebraic quantum field theory. It is shown that for all positive temperatures
there are fermionic solutions only if the coupling constant is $\lambda =
\sqrt{2(2n+1)\pi}, \, n\in \bf N$, otherwise solutions are anyons.
Different anyons (which are uncountably many) live in orthogonal spaces, so the
whole Hilbert space becomes non-separable and in
each of its sectors a different Urgleichung holds.
This feature certainly cannot be seen by any power expansion in $\lambda$.
Moreover, if the statistic parameter is tied to the coupling constant it is
clear that such an expansion is doomed to failure and will never reveal the
true structure of the theory.

On the basis of the model in question, it is not possible to decide whether
fermions or bosons are more fundamental since dressed fermions can be
constructed either from bare fermions or directly from the current algebra.

\vspace{8pt}
\begin{center}
{\it Invited talk at the \\
XI International Conference \\
PROBLEMS OF QUANTUM FIELD THEORY \\[6pt]
In memory of \\
D.I. Blokhintsev\\[8pt]
July 1998, Dubna, Russia}
\end{center}
\vspace{10pt}
\vfill
{\footnotesize

$^\star$ Work supported in part by ``Fonds zur F\"orderung der
wissenschaftlichen Forschung in \"Osterreich" under grant P11287--PHY; 

$^\ast$ Permanent address: Institute for Nuclear Research and Nuclear Energy,
Bulgarian Academy of Sciences, Boul.Tzarigradsko Chaussee 72, 1784 Sofia,
Bulgaria

$^\sharp$ E--mail address: ilieva@pap.univie.ac.at}

\end{titlepage}

\section{Introduction}
After T.D. Lee had constructed a model of a soluble QFT \cite{TDL} many people
tried to find other examples; but to solve a nontrivial relativistic QFT seemed
out of the question. The idea that Bethe's ansatz \cite{B} could be
successfully used to solve also Heisenberg's ``Urgleichung" \cite{Hei} reduced 
to one
space one time dimension then led to a soluble relativistic field theory -- the
Thirring model \cite{WT}. During the years, this model has not only been 
extensively studied but has also been actively used for analysis, testing
and illustration of various phenomena in two--dimensional field theories. 

It is not our purpose to review the enormous literature on the subject but we
rather focus on the very starting point -- Heisenberg's Urgleichung.
With no bosons present in it at all, it represents
the ultimate version of the opinion  that fermions should enter the basic 
formalism of the fundamental theory of elementary particles that is usually 
taken for granted.
 
The opposite point of view, namely that a theory including only
observable fields, necessarily uncharged bosons, is capable of describing
evolution and symmetries of a physical system, being the kernel of algebraic
approach to QFT \cite{H}, also enjoys an enthusiastic support. 
As we will see, there is no possibility to judge this matter on the basis of
the model in question, since both  formulations can be equally well used to
construct the physically relevant objects -- the dressed fermions. 

In any case, before claiming that an ``Urgleichung" of the type
\beq
\not\!\partial\psi(x) = \lambda\psi(x)\bar\psi(x)\psi(x)
\eeq
determines the whole Universe one should see whether it determines anything
mathematically and it is our aim in the present note to discuss the elements
needed to make its solution well defined. In fact we shall first consider only
one chiral component and we shall restrict ourselves to the 
two--dimensional spacetime, so that this component depends only on one light
cone coordinate. Also the bose--fermi duality takes 
place there and we want to make use of it. This phenomenon amounts to the fact 
that in certain models formal functions of fermi fields can be written that 
have vacuum expectation values and statistics of bosons and vice versa, the 
equivalence being understood within perturbation theory.

The bose--fermi duality is actually well established when the construction of
bosons out of fermions is considered.
The problem of rigorous definitions of operator--valued distributions and
eventually operators having the basic properties of fermions by taking
functions of bosonic fields is rather more delicate. On the level of operator
valued distributions solutions have been given by Dell'Antonio et al.\cite{DA}
and Mandelstam \cite{M} and on the level of operators in a Hilbert space --- by
Carey and collaborators \cite{CR, CHB} and in a Krein space by Acerbi, Morchio
and Strocchi \cite{AMS}.

Thus our goal is to give a precise meaning to the following three ingredients
\beq
\begin{array}{llcl}
(a) & [\,\psi^\ast(x), \psi(x')\,]_+ = \delta(x\!-\!x'), \qquad 
[\,\psi(x), \psi(x')\,]_+ = 0 & \qquad & {\rm CAR} \\[7pt]
(b) & \frac{1}{i}\frac{d}{dx}\psi(x) = \lambda j(x)\psi(x) & 
\qquad & {\rm Urgleichung}\\[7pt]
(c) & j(x) = \psi^\ast(x)\psi(x) & \qquad & {\rm Current}
\end{array}
\eeq

Eq.(1.2b) involves (derivatives of) objects which are according to (1.2a)
rather discontinuous. Therefore it is expedient to pass right away to the level of
operators in Hilbert space since the variety of topologies there provides a
better control over the
limiting procedures. In general norm convergence can hardly be hoped for but we
have to strive at least for strong convergence such that the limit of the
product is the product of the limits. With $\psi_f = \int_{-\infty}^{\infty}dx
f(x)\psi(x)$, (1.2a) becomes
\beq
[\,\psi^\ast_f, \psi_g\,]_+ = \langle f\vert g\rangle
\eeq
for $f\in L^2(\bf R)$ and $\langle .\vert .\rangle$ the scalar product in
$L^2(\bf R)$. This shows that $\psi_f$'s are bounded and form the
$C^\ast$--algebra CAR. There the translations $\,x\ra x+t\,$ give an automorphism
$\tau_t$ and we shall use the corresponding KMS--states $\omega_\beta$ and the
associated representation $\pi_\beta$ to extend CAR. Though there $j = \infty$,
one can give a meaning to $j$ as a strong limit in $\Ha_\beta$ by smearing
$\psi(x)$ over a region $\ve$ to $\psi_\ve(x)$ and define
$$
j_f = \int dx f(x) \lim_{\ve\to 0}\left(\psi^*_\ve(x)\psi_\ve(x) -
\omega_\beta(\psi^*_\ve(x)\psi_\ve(x))\right), \qquad f:{\bf R}\to {\bf R}
$$
These limits exist in the strong resolvent sense and define self--adjoint
operators which determine with
\beq
e^{ij_f}e^{ij_g} = e^{\frac{i}{8\pi}\int dx(f(x)g'(x)-f'(x)g(x))}e^{ij_{f+g}}
\eeq
the current algebra $\A_c$. Its Weyl structure is the same for all $\beta > 0$ and
$\omega_\beta$ extends to $\A_c$.

To construct the interacting fermions which on the level of distributions look
like 
$$
\Psi(x) = Z \,e^{i\lambda\int_{-\infty}^{x}dx'j(x')} \st{?}{=}
\lim_{\ve \dar 0} \lim_{R \ra \infty} \Psi_{\ve, R}(x)
$$ 
(with some renormalization constant $Z$) poses both infrared ($R \ra \infty$) 
and ultraviolet ($\ve \ra 0$) problems.
So an extension of $\pi(\A_c)''$ is needed to accomodate such a kind of objects.

There are two equivalent ways of handling the infrared problem. Since the
automorphism generated by the unitaries $\Psi_{\ve, R}(x)$ converges to a
limit $\gamma$ for $R\to\infty$, one can form with it the crossed product 
$ \bar\A_c = \A_c \,\st{\gamma}{\bowtie}\,\rm Z$, so that in $\bar\A_c$ there
are unitaries with the properties which the limit should have \cite{IN, IT}. On 
the other hand, the symplectic form in (1.4) and the state $\omega_\beta$ can be 
defined for the limiting element $\Psi_\ve(x)$ and this we shall do in 
what follows.

In  any case $\bar\Ha_\beta$ assumes a sectorial structure, the subspaces  
$\A_c\,\dprod_{i=1}^{n} \Psi_\ve(x_i)\vert\Omega\rangle\,$ for different
$n$ are orthogonal and thus may be called $n$--fold charged sectors. The
$\Psi_\ve(x)$'s have the property that for $\vert x_i - x_j\vert > 2\ve$
they obey anyon statistics with parameter $\lambda^2$ and an Urgleichung (1.2b)
where $j(x)$ is averaged over a region of lenght $\ve$ below $x$. 

Then, by removing the ultraviolet cut--off 
the sectors abound and the subspaces
$\,\A_c\Psi(x)\vert\Omega\rangle\,$ become orthogonal for different $x$, so
$\bar\Ha_\beta$ becomes non--separable. To get canonical fields of the type
(1.3) one has to combine $\ve\dar 0$ with a field renormalization
$\Psi_\ve \to \ve^{-1/2}\Psi_\ve$ such that
$$ 
\lim_{\ve\dar 0}\ve^{-1/2}\int dx f(x)\Psi_\ve(x) = \Psi_f 
$$
converge strongly in $\bar\Ha_\beta$ and satisfy (1.2b) in sense of
distributions.

However, the objects so constructed are in general anyons and only for
particular values of the coupling constant, $\lambda = 
\sqrt{2(2n+1)\pi},\,\, n\in\bf N$, they are fermions, so that the coupling
constant is tied to the statistic parameter. Thus we find that there is
indeed some magic about the Urgleichung inasmuch as on the quantum level it
allows fermionic solutions by this construction only for isolated values of the coupling
constant $\lambda$ whereas classically $\Psi(x) = Z \,e^{i\lambda\int_
{-\infty}^{x}dx'j(x')}$ solves (1.2b) for any $\lambda$. This feature can 
certainly not be seen by any power expansion in $\lambda$.

By a {\it symmetry} $\,\alpha\,$ of a physical system an automorphism of the 
algebra $\,\A\,$ which describes it is understood. The algebraic chain of 
inclusions we construct gives an example of a {\it symmetry destruction}, that 
is, for a given extension $\,\B\,$ of the algebra $\,\A\,$, $\,\B \supset \A$,
$\,\not\hspace{-1mm}\exists \beta \in \mbox{Aut } \B: \,\, \beta\vert_\A = 
\alpha$ for some $\,\alpha \in \mbox{Aut }\A$. This
phenomenon is related to the spontaneous collapse of a symmetry \cite{BO} and 
in contrast to the spontaneous symmetry breaking \cite{NT}, it cannot occur in 
a finite--dimensional Hilbert space.
\vspace{0.6cm}

\section{Bosons out of fermions: the CAR-algebra, its KMS-states and associated 
v. Neumann algebras}
Let us consider the  C*-algebra $\A^l$ formed by the bounded operators
\beq
\psi_f = \int_{-\infty}^\infty dx \psi(x) f(x) =
\int_{-\infty}^\infty \frac{dp}{2\pi} \wt \psi(p) \wt f(p), \qquad
\wt f(p) = \int_{-\infty}^\infty dx \; e^{ipx} f(x)
\eeq
with $\psi(x)$, $x \in {\bf R}$, being operator-valued distributions
which satisfy
\beq
[\psi^*(x),\psi(x')]_+ = \delta(x\!-\!x'),
\eeq
so, describing the left movers (we have asigned a superscriped to the relevant
quantities, $x$ stands for $x-t$) and $f \in L^2({\bf R})$. This algebra is 
characterized by
\beq
[\psi^*_f,\psi_g]_+ = \langle f|g\rangle = \int dx f^*(x) g(x).
\eeq
Translation $\tau_t$ define an automorphism of $\,\A^l\,$
\beq
\tau_t \psi_f = \psi_{f_t}, \quad
f_t(x) = f(x-t).
\eeq
$\A^l\,$ inherits the norm from $L^2({\bf R})$ such that $\,\tau_t\,$ is
(pointwise) normcontinuous in $\,t\,$ and even normdifferentiable for the
dense set of $f$'s for which
$$
\lim_{\delta \dar 0} \frac{f(x+\delta)-f(x)}{\delta} = f'(x)
$$
exists in $L^2({\bf R})$
\beq
\left.\frac{d}{dt} \; \tau_t \psi_f \right|_{t=0} = - \psi_{f'}.
\eeq
The $\tau$-KMS-states over $\A^l\,$ are given by
\beq
\omega_\beta(\psi^*_f \psi_g) = 
\int_{-\infty}^\infty \frac{dp}{2\pi} \frac{\wt f^*(p) \wt g(p)}
{1 + e^{\beta p}} =
\sum_{n=-\infty}^\infty \frac{(-1)^n}{2\pi}
\int \frac{dx dx' f^*(x) g(x')}{i(x-x') - n\beta + \ve}, \qquad
\ve \dar 0,
\eeq
$$
\omega_\beta(\psi_g \psi^*_f) = 
\omega_\beta(\psi^*_f \tau_{i\beta} \psi_g).
$$
With each $\omega_\beta$ are associated a representation $\pi_\beta$ with
cyclic vector $|\Omega\rangle\,$, $\,\omega_\beta(a) = \langle \Omega|a|\Omega
\rangle$ in $\,\Ha_\beta = \ol{\A^l|\Omega\rangle}\,$ and a v.~Neumann algebra
$\,\pi_\beta(\A^l)''$. It contains the current algebra $\A^l_c$ which gives
the formal expression $j(x) = \psi^*(x) \psi(x)$ a precise meaning.

To show this, let us recall two lemmas (for the proofs see \cite{IT}) which make
the whole construction transparent:
\paragraph{Lemma} (2.7) \\
If the kernel $K(k,k'): {\bf R}^2 \ra C$ is as operator $\geq 0$ and
trace class $(K(k,k) \in L^1({\bf R}))$, then $\forall \; \beta \in
{\bf R}^+$
\beqan
\lim_{M \ra \pm \infty} B_M &:=& \lim_{M \ra \pm \infty}
\frac{1}{(2\pi)^2}\int dk dk' K(k,k') \wt \psi^*(k+M) \wt \psi(k'+M) = \\
&=& \frac{1}{(2\pi)^2}\int dk dk' \lim_{M \ra \pm \infty} K(k,k') \omega_\beta
(\wt \psi^*(k+M) \wt \psi(k'+M)) = \\
&=& \left\{\ba{cl} \frac{1}{2\pi}\int dk \; K(k,k) & \mbox{ for } M \ra +\infty \\
0 & \mbox{ for } M \ra - \infty \ea \right.
\eeqan
in the strong sense in $\Ha_\beta$. \\[3pt]

However, if $\int |K|^2$ keeps increasing with $M$, then 
$B_M - \langle B_M\rangle$ may nevertheless tend to an (unbounded) operator.

\paragraph{Lemma} (2.8) \\
If
$$
B_M = \frac{1}{(2\pi)^2}\int dk dk' \wt f(k-k') \Theta(M-|k|) 
\Theta(M-|k'|) \wt \psi^*(k) \psi(k')
$$
with $\wt f$ decreasing faster than an exponential and being the Fourier 
transform 
of a positive function, then the difference $\,B_M - \omega_\beta(B_M)\,$ is a 
strong Cauchy sequence $\,\ra 0\,$ for $\,M \ra \infty\,$ on a dense domain on 
$\Ha_\beta$.

\paragraph{Remarks} (2.9)
\begin{enumerate}
\item (2.7) substantiates the feeling that for $k > 0$ most levels are
empty and for $k < 0$ most are full.
\item $B_M$ is a positive operator and by diagonalizing $K$ one sees
$$
\|B_M\| = \|K\|_1 = \frac{1}{2\pi}\int dk\; K(k,k).
$$
\item As just mentioned, $\,\|B_M\| < 2M \wt f(0)\,$ and $f(x) \geq 0$ is not 
a serious restriction since any function is a linear combination of positive
functions.
\item Since the limit $j_f$ is unbounded the convergence is not on all
of $\Ha_\beta$, however since for the limit $j_f$ holds $\tau_{i\beta}j_f = 
j_{e^{\beta p} f}$,  the dense domain is invariant under $j_f$. Thus we have
strong resolvent convergence which means that bounded functions of
$B_M$ converge strongly. Also the commutator of the limit is the limit
of the commutators.
\end{enumerate}
\vspace{0.4cm}

Thus we conclude that the limit exists and is selfadjoint on a suitable domain.
We shall write it formally
$$
j_f = \int_{-\infty}^\infty \frac{dk dk'}{(2\pi)^2} \; \wt f(k-k')
: \wt \psi(k)^* \wt \psi(k'):  \eqno(2.10)
$$

Next we show that the currents so defined satisfy the CCR with a suitable symplectic form
$\sigma$ \cite{J, Schwinger}.
\paragraph{Theorem} (2.11)
$$
[j_f,j_g] = i \sigma(f,g) =
\int_{-\infty}^\infty \frac{dp}{(2\pi)^2} \; p \wt f(p) \wt g(-p) =
\frac{i}{4\pi} \int_{-\infty}^\infty dx(f'(x)g(x) - f(x)g'(x)).
$$
\paragraph{Proof:} For the distributions $\wt \psi(k)$ we get algebraically
$$
[\wt\psi^*(k) \wt\psi(k'),\wt\psi^*(q)\wt\psi(q')] =
2\pi\left[\wt\psi^*(k)\wt\psi(q')\delta(q\!-\!k') -
\wt\psi^*(q)\wt\psi(k') \delta(k\!-\!q')\right]
$$
and for the operators after some change of variables
$$
\frac{1}{(2\pi)^3}\int dkdpdp' \wt f(p) \wt g(p') \wt\psi^*(p+p'+k) \wt\psi(k)
\Theta(M-|k|) \Theta(M-|p+p'+k|) \cdot
$$
$$
\cdot \left[\Theta(M-|p'+k|) - \Theta(M-|p+k|)\right].
$$
For fixed $p$ and $p'$ and $M \ra \infty$ we see that the allowed region
for $k$ is contained in $(M-|p|-|p'|,M)$ and
$(-M,-M+|p|+|p'|)$. Upon $k \ra k \pm M$ we are in the situation of (2.7),
thus we see that the commutator of the currents (2.10)
is bounded uniformly in $M$ if $\wt f$ and 
$\wt g$ decay faster than exponentials and converges to the expectation
value. This gives finally
$$
\int_{-\infty}^\infty \frac{dp}{(2\pi)^2} \; \wt f(p) \wt g(-p)
\int dk \; \Theta(M-|k|) \left[\Theta(M - |k-p|) - \Theta(M - |k+p|)\right]
\frac{1}{1 + e^{\beta k}}
$$
$$
\st{M \ra \infty}{\longrightarrow} \int_{-\infty}^\infty \frac{dp}{(2\pi)^2}
\; p \wt f(p) \wt g(-p).
$$

\paragraph{Remarks} (2.12)
\begin{enumerate}
\item Since the $j_f$'s satisfy the CCR they cannot be bounded and it is
better to write (2.11) in the Weyl form for the associated unitaries
$$
e^{ij_f} \; e^{ij_g} = e^{\frac{i}{2} \sigma(g,f)} \; e^{ij_{f+g}} =
e^{i\sigma(g,f)} \; e^{ij_g} \; e^{ij_f}.
$$ 
\item The currents $j_f$ are selfadjoint, so the unitaries $\,e^{i\alpha j_f}$ 
generate 1--parameter groups --- the local gauge transformations
$$
e^{-i\alpha j_f} \; \psi_g \; e^{i\alpha j_f} = \psi_{e^{i\alpha f}g}.
$$
\item The state $\,\omega_\beta\,$ can be extended to $\,\bar \omega_\beta\,$
over $\,\pi_\beta(\A^l)''$ and $\tau_t$ to $\bar \tau_t,\,\,
\bar \tau_t \in \mbox{Aut }\pi_\beta(\A^l)''$ with $\bar \tau_t \,j_f = j_{f_t}$.
Furthermore $\,\bar \omega_\beta\,$ is $\,\bar \tau$--KMS and is calculated to 
be (\cite{IT}, see also \cite{BY})
$$
\bar \omega_\beta(e^{ij_f}) = \exp \left[ -\frac{1}{2} \int_{-\infty}^\infty
\frac{dp}{(2\pi)^2} \frac{p}{1 - e^{-\beta p}} |\wt f(p)|^2  \right].
$$
\item A physically important symmetry of the algebra $\,\A^l\,$, the parity $P$,
$$ 
P \in \mbox{Aut }\A^l, \quad  P\psi_f = \psi_{Pf}, \quad P f(x) = f(-x)
$$
is destroyed in $\pi_\beta$, since 
$$
[j(x),j(x')] = -\,\frac{i}{2\pi}\,\delta'(x\!-\!x')
$$
is not invariant under $j(x) \ra j(-x)$. 
Thus $P \notin \mbox{Aut }\pi_\beta(\A^l)''$ and $\,\bar\omega_\beta$ is not
P--invariant. 
\item The extended shift automorphism $\,\bar \tau_t\,$ is not only strongly 
continuous but for
suitable $f$'s also differentiable in $\,t\,$ (strongly on a dense set in 
$\Ha_\beta$)
$$
\frac{1}{i} \frac{d}{dt} \bar \tau_t e^{ij_f} = 
\left[ j_{f'_t} + \frac{1}{2} \sigma(f_t,f'_t)  \right] e^{ij_{f_t}} =
e^{ij_{f_t}} \left[j_{f'_t} - \frac{1}{2} \sigma(f_t,f'_t) \right] =
\frac{1}{2} \left[ j_{f'_t} \; e^{ij_{f_t}} + e^{ij_{f_t}} j_{f'_t} \right].
$$
\item The symplectic structure is formally independent on $\beta$ \cite{HG}, 
however for $\beta < 0$ it changes its sign, $\sigma \to -\sigma$, and for 
$\beta = 0$ (the tracial state) it becomes zero.
\end{enumerate}

Thus starting from a CAR-algebra $\A^l$, we identified in $\pi_\beta(\A^l)''$
bosonic fields -- the currents, which satisfy CCR's. The crucial ingredient
needed was the appropriately chosen state. Here we have used the KMS--state
(which is unique for the CAR algebra). Another possibility would be to
introduce the Dirac vacuum (filling all negative energy levels in the Dirac
sea). This is  what has been done in the thirties \cite{J, BNN}, in order
to achieve stability for a fermion system, and recovered later by Mattis and
Lieb \cite{ML} in the context of the Luttinger model. Thus as an additional 
effect the appearance of an anomalous term in the current commutator (later 
called Schwinger term) had been discovered that actually enables bosonization 
of these two--dimensional models.

\vspace{0.6cm}

\section{Extensions of $\A_c$: fermions out of bosons}
So far $\A^l_c$ was defined for $j_f$'s with $\, f \in C_0^\infty$, for
instance. The algebraic structure is determined by the symplectic form 
$\sigma(f,g)$ (2.11) which 
is actually well defined also for the 
Sobolev space, $\sigma(f,g) \ra \sigma(\bar f, \bar g), \, \bar f, \bar g \in
H_1, \, H_1 = \{f : f,f' \in L^2\}\,$. Also $\bar \omega_\beta$
can be extended to $H_1$, since $\,\bar \omega_\beta(e^{ij_{\bar f}}) > 0\,$
for $\,\bar f \in H_1$.
The anticommuting operators we are looking for are of the form
$e^{ij_f}$, with $f(x) = 2\pi \Theta(x_0\!-\!x) \not\in H_1$. Still one can give
$\sigma(f,g)$ a meaning for such an $f$. However, the corresponding state 
$\omega_\beta$ exhibits singular behaviour for both $p \ra 0$ and 
$p \ra \infty$, so that 
$$
\omega_\beta(e^{ij_\Theta}) = \exp \left[-\frac{1}{2} \int_{-\infty}^\infty
\frac{dp}{p(1 - e^{-\beta p})} \right] = 0
$$
and thus such an operator would act in $\Ha_\beta$ as zero. Therefore 
an approximation of $\Theta$ by functions from $H_1$ would result in unitaries
that converge weakly to zero.

This situation can be visualized by the following 
\paragraph{Example} (3.1) \\
Consider the $H_1$--function $\,\Phi_{\delta,\ve}(x)$,
$$
\Phi_{\delta,\ve}(x) := \vp_\ve(x) - \vp_\ve(x + \delta) \in H_1, 
$$
with
$$
\vp_\ve(x) := \left\{\ba{cl} 1 & \mbox{ for } x \leq -\ve \\
-x/\ve & \mbox{ for } - \ve \leq x \leq 0 \\
0 & \mbox{ for } x \geq 0 \ea \right.
$$
as an approximation to the step function,
$$
\lim_{\delta \ra \infty \atop \ve \ra 0} \Phi_{\delta,\ve}(x) =
\Theta(x).
$$
Then
$$
\wt \Phi_{\delta,\ve}(p) = \frac{1 - e^{ip\ve}}{\ve p^2} (1 - e^{ip\delta})
$$
and
$$
\|\Phi_{\delta,\ve}\|^2_\beta = \int_{-\infty}^\infty \frac{dp}{2\pi} 
\frac{p}{1 - e^{-\beta p}}
| \wt \Phi(p)|^2 =
16 \int_{-\infty}^\infty \frac{dp}{2\pi} \frac{p}{1 - e^{-\beta p}}
\frac{\sin^2 p \ve/2}{\ve^2 p^4} \sin^2 p \delta/2 , 
$$
$$
\|\Phi\|^2_\beta \geq c \int_0^{1/\delta} dp \; \delta^2 = c \delta
$$
for $\beta/\delta$, $\ve/\delta \ll 1$ and $c$ a constant. Thus for
$\delta \ra \infty\,$, $\,\|\Phi_\delta\|_\beta \ra \infty$. Also $\,
\|\Phi_\delta - f\|_\beta \ra \infty$ since 
$$
\|\Phi_\delta - f\|_\beta \geq \|\Phi_\delta\|_\beta - \|f\|_\beta \ra \infty
\qquad \forall \; \|f\|_\beta < \infty
$$
and thus
$$
|\langle \Omega|e^{-ij_f} \; e^{ij_{\Phi_\delta}}|\Omega\rangle| =
e^{-\frac{1}{2} \|\Phi_\delta - f\|_\beta^2} \ra 0.
$$
But $e^{ij_f}|\Omega\rangle$, $\|f\|_\beta < \infty$, is total in 
$\Ha_\beta$ and thus $e^{ij_{\Phi_\delta}}|\Omega\rangle$ and therefore
$e^{ij_{\Phi_\delta}}$ goes weakly to zero. 

However the automorphism
$$
e^{ij_f} \ra e^{-ij_{\Phi_\delta}} e^{ij_f} e^{ij_{\Phi_\delta}} =
e^{i \sigma(\Phi_\delta,f)} e^{ij_f}
$$
converges since
$$
\sigma(f,\Phi_\delta) = - \frac{1}{2\pi\ve} \left( \int_{-\ve}^0 -
\int_{-\ve - \delta}^{-\delta}\right) dx \; f(x) 
\st{\delta \ra \infty}{\longrightarrow}
- \frac{1}{2\pi\ve} \int_{-\ve}^0 dx \; f(x)
\st{\ve\ra 0}{\longrightarrow}
-\frac{1}{2\pi}\,f(0).
$$
The divergence of $\|\Phi_{\delta,\ve}\|$ is related to the well--known
infrared problem of the massless scalar field in $(1\!+\!1)$ dimensions and
various remedies have been proposed \cite{S}. We take it as a sign
that one should enlarge $\,\A^l_c\,$ to some $\,\bar \A^l_c\,$ and work in the 
Hilbert space $\bar \Ha\,$ generated by $\,\bar \A^l_c\,$ on the natural extension 
of the state. Thus we add to $\,\A^l_c\,$ the idealized element 
$\,e^{i 2\pi j_{\vp_\ve}} = U_\pi\,$
and keep $\sigma$ and $\omega_\beta$ as before.
Equivalently we take the automorphism $\,\gamma\,$ generated by $\,U_\pi\,$ and
consider the crossed product $\,\bar \A^l_c = \A^l_c \st{\gamma}{\bowtie} 
\bf Z\,$. There is a natural extension $\,\bar \omega\,$ to $\,\bar \A^l_c\,$ 
and a natural isomorphism of $\,\bar \Ha\,$ and $\,\bar \A^l_c|\bar \Omega\rangle$.
Here $\bar \Ha$ is the countable orthogonal sum of sectors with $n$
particles created by $\,U_\pi$. Thus,
$$
\langle \Omega|e^{ij_f} U_\pi|\Omega\rangle = 0 \eqno(3.2)
$$
means that $U_\pi$ leads to the one--particle sector, 
in general
$$
\langle \Omega|U_\pi^{*n} \, e^{ij_f} U_\pi^m|\Omega\rangle =
\delta_{nm}\,\omega_\beta(\gamma^n\,e^{ij_f}).
$$ 
The quasifree automorphisms on $\,\A^l_c$ (e.g. $\tau_t$) 
can be naturally extended to $\,\bar \A^l_c\,\,$, $\tau_t \,U_\pi~=~
e^{i\pi j_{\vp_{\ve,t}}}$, $\,\vp_{\ve,t}(x) = \vp_\ve(x\!+\!t)\,$  and since
$\vp_\ve - \vp_{\ve,t} \in H_1$ $\,\forall \; t$, this does not lead out
of $\bar \A^l_c$. 

$U_\pi$ has some features of a fermionic field since
$$
\sigma(\vp_\ve,\tau_t \vp_\ve) = - \sigma(\vp_\ve,\tau_{-t} \vp_\ve) =
\frac{1}{4\pi}\left\{\ba{cl} 1 & \mbox{ for } t > \ve \\
\frac{2t}{\ve} - \frac{t^2}{\ve^2} & \mbox{ for } 0 \leq t \leq \ve \ea \right. . \eqno(3.3)
$$
More generally we could define $U_\alpha = e^{i \sqrt{2\pi\alpha} j_{\vp_\ve}}$
and get from (3.3) with 
$$
\sgn(t) = \Theta(x) - \Theta(-x) = \left\{
\begin{array}{cl} 1 & {\rm for} \qquad t > 0 \\
0 & {\rm for} \qquad t = 0 \\
-1 & {\rm for} \qquad t <  0.
\end{array} \right.
$$
\paragraph{Proposition} (3.4)
\beqan
U_\alpha \tau_t U_\alpha &=& \tau_t(U_\alpha) U_\alpha \,
e^{i\,\alpha\,\sgn(t)/2}, \\
U^*_\alpha \tau_t U_\alpha &=& \tau_t(U_\alpha) U^*_\alpha \,
e^{i\,\alpha\,\sgn(t)/2} \quad \forall \; |t| > \ve.
\eeqan

\paragraph{Remark} (3.5) \\
We note a striking difference between the general case of anyon statistics 
and the two particular cases --- Bose ($\alpha = 2\cdot 2n\pi$) or Fermi 
($\alpha = 2(2n+1)\pi$) 
statistics. Only in the latter two cases parity $P$ (2.12:4) is an
automorphism of the extended algebra generated through $U_\alpha$. Thus $P$
which was destroyed in $\A^l_c$ is now recovered for two subalgebras.\\[2pt]

The particle sectors are orthogonal in any case
$$
\langle \Omega| U^*{}^n_\alpha \,e^{ij_f} U^m_\alpha|\Omega\rangle =
0 \quad \forall \; n \neq m, \; f \in H_1.
$$
Furthermore, sectors with different statistics are orthogonal
$\langle\Omega\vert U^*_\alpha U_{\alpha'}\vert\Omega\rangle = 0, \, \alpha
\not= \alpha'$, thus if we adjoin $U_\alpha, \, \forall \alpha\in{\bf R_+}$,
$\,\bar\Ha_\beta$ becomes nonseparable.
 
Next we want to get rid of the ultraviolet cut--off and let $\ve$ go to
zero. Proceeding the same way we can extend $\sigma$ and $\tau_t$ but 
keeping $\omega$ the sectors abound. The reason is that
$\vp_\ve \st{\ve \ra 0}{\longrightarrow} \Theta(x)$ and
$$
\|\Theta - \Theta_t\|^2 = \int_{-\infty}^\infty 
\frac{dp \; p}{1 - e^{-\beta p}} \frac{|1 - e^{itp}|^2}{p^2}
$$
is finite near $p = 0$ but diverges logarithmically for $p \ra \infty$.
This means that $e^{ij_f} e^{ij_\Theta}|\Omega\rangle$, $f \in H_1$
gives a sector where one of these particles (fermions, bosons or anyons)
is at the point $x = 0$ and is orthogonal to $e^{ij_f} e^{ij_{\Theta_t}}
| \Omega\rangle\,$ $\forall \; t \not= 0$. Thus the total Hilbert space is not
separable and the shift $\tau_t$ is not even weakly continuous, so
there is no chance to make sense of $\frac{d}{dt} \tau_t e^{ij_\Theta}$.

So far, only one chiral component has been considered. When both chiralities
are present, no significant changes arise in the construction described. The
only point demanding for some care is the anticommutativity between left- and
right- moving fermions, which asks for an even larger extension of the current
algebra by extending its test--functions space. 

So, the Weyl algebra $\,\A_c = \A_c^r \otimes \A_c^l \,$ is now generated by the
unitaries
$$
W(f_r, f_l) = e^{i\int\left( f_r(x)j_r(x) + f_l(x)j_l(x)\right)dx},
$$
with $\sigma(f_l, g_l)$ given by (2.11) and $\sigma(f_r, g_r) = 
-\sigma(f_l, g_l)$. The minimal extension of $\,\A_c\,$ is then obtained by
adding two idealized elements, 
$$
U_\pi^l := W(c_l, \,2\pi(1-\vp_\ve)) \quad \mbox{and }\quad
U_\pi^r := W(2\pi\vp_\ve, \,c_r)
$$
$$ 
c_l - c_r = (2k + 1)\pi, \quad k \in {\bf Z} \qquad
(\mbox{e.g. } c_r = \pi/2 = -c_l)
$$
They generate for $\ve \ra 0$ (not inner) automorphisms of $\,\A_c\,$
 $$
 U_\pi^{r(l)} : \quad \gamma_{r(l)}\,W(f_r, f_l) = e^{if_{r(l)}(0)}
 \,e^{\frac{i}{4}\int_{-\infty}^\infty f'_{r(l)}(y)dy}\,W(f_r,f_l)
 $$
 in which for the two subalgebras $\,\{W(\bar f, \bar f)\}\,$ and $\,\{W(\bar f,
 -\bar f)\},\,\, \bar f \in \D_o\subset \C_o^\infty,\,$ the
 vector and axial gauge transformations can easily be traced back.
 
 In addition to the obvious replacement of (3.4),  also the following relation
 holds
 $$
 U_\pi^{r(*)}\, U_\pi^l = - U_\pi^l \, U_\pi^{r(*)}.
 $$
 
 Thus we can identify the chiral components of the fermion field with the so
 constructed unitaries
 $$
 \psi_r(x) = exp\,\{i2\pi\int^{\infty}_{-\infty} \vp_\ve(y-x)j_r(x')dx' 
 \pm i\frac{\pi}{2}\int_{-\infty}^\infty j_l(x')dx'\} 
 $$
 $$
 \psi_l(x) = exp\,\{\mp i\frac{\pi}{2}\int_{-\infty}^\infty j_r(x')dx' 
 + i2\pi\int^{\infty}_{-\infty} \vp_\ve(x-y) j_l(x')dx'\} \eqno(3.6)
 $$
 
 In general, we could define an extension of the algebra $\,\A_c\,$ through the
 abstract elements $\,U_\alpha^{r(l)}\,$
 \beqan
 U_{\alpha}^r &:=& W(\sqrt{2\pi\alpha}\, \vp_\ve, \,
 \frac{1}{2}\sqrt{\frac{\pi\alpha}{2}}) \\ 
 U_{\alpha}^l &:=& W(-\frac{1}{2}\sqrt{\frac{\pi\alpha}{2}}, 
 \,\sqrt{2\pi\alpha}\, (1-\vp_\ve)) \no
 \eeqan
 
 Propositon (3.4) extends also for the non--chiral model generalization.
 As expected, admitting arbitrary values for $\alpha$, we get a very rich field
 structure where definite statistic behaviour is preserved only within a given
 field class (fixed value of $\alpha$), so that even ``different" fermions
 (with different ``2 x odd" values of $\alpha$) do not anticommute but instead 
 follow the general fractional statistics law.
\vspace{0.6cm}

\section{Anyon fields in $\pi_{\bar \omega}(\bar \A_c)''$}
Next we shall use another ultraviolet limit to construct local fields
which obey some anyon statistics. Of course quantities like
$$
[\Psi^*(x),\Psi(x')]_\alpha := 
\Psi^*(x) \Psi(x') e^{i \frac{2\pi-\alpha}{4} \sgn(x'-x)} 
+ \Psi(x') \Psi^*(x) e^{-i \frac{2\pi-\alpha}{4} \sgn(x'-x)} =
\delta(x\!-\!x')
$$
will only be operator valued distributions and have to be smeared to give
operators. Furthermore in this limit the unitaries we used so far have to 
be renormalized so that $\delta(x\!-\!x')$ gets a factor 1 in front. A candidate
for $\Psi(x)$ will be ($\alpha \in (0, 4\pi)$)
$$
\Psi(x) := 
\lim_{\ve \ra 0} n(\ve) \exp \left[ i \sqrt{2\pi\alpha} \int_{-\infty}^\infty
dy \; \vp_\ve(x-y) j(y) \right] 
$$
with $\vp_\ve$ from (3.1) and $n(\ve)$ a suitably chosen normalization.
With the shorthand $\vp_{\ve,x}(y) = \vp_\ve(x-y)$ we can write
\beqan
\Psi^*_\ve(x) \Psi_\ve(x') &=& 
\exp \left\{ i \,2\pi\alpha \,\sigma(\vp_{\ve,x},\vp_{\ve,x'}) \right\}
\exp \left\{ i \sqrt{2\pi\alpha} \, j_{\vp_{\ve,x'} - \vp_{\ve,x}}\right\}, \\
\Psi_\ve(x') \Psi^*_\ve(x) &=&
\exp \left\{- i \,2\pi\alpha \,\sigma(\vp_{\ve,x},\vp_{\ve,x'}) \right\}
\exp \left\{ i \,\sqrt{2\pi\alpha} \,j_{\vp_{\ve,x'} - \vp_{\ve,x}}\right\}.
\eeqan
We had in (3.3)
$$
4\pi\sigma(\vp_{\ve,x},\vp_{\ve,x'}) = \sgn(x-x')
\left\{ \Theta(|x-x'| - \ve) + \Theta(\ve - |x-x'|)
\frac{(x - x')^2}{\ve^2} \right\} 
$$
$$
=: \sgn(x-x') D_\ve(x-x')
$$
and thus
$$
[\Psi^*_\ve(x),\Psi_\ve(x')]_\alpha = 2n(\ve)^2 \cos \left[\sgn(x-x')
\left( \frac{\pi}{2} - \frac{\alpha}{4}(1 - D_\ve(x-x'))\right)\right]
\exp \left[ i \alpha j_{\vp_{\ve,x'} - \vp_{\ve,x}}\right].
$$
Note that for $|x-x'| \geq \ve\,$ the argument of the $\cos$ becomes 
$ \pm \pi/2$, so the $\alpha$--commutator vanishes, in agreement with (3.4). 
To manufacture a $\delta$-function for $|x-x'|
\leq \ve$ we note that $cos(...) > 0$ and $\omega_\beta(e^{i\alpha j}) > 0$, so 
we have to choose $n(\ve)$ such that
$$
2 n^2(\ve)\ve \int_{-1}^1 d\delta \cos \left( \frac{\pi}{2} -
\frac{\alpha}{4}(1 - \delta^2) \right) \cdot \omega_\beta
\left(\exp \left[ i \alpha j_{\vp_{\ve,x-\ve\delta} - \vp_{\ve,x}}\right]\right)
= 1
$$
and to verify that for $\ve \dar 0$ $\,[\;]_\alpha$ converges 
strongly to a $c$-number. For the latter to be finite we have to smear $\Psi(x)$ with
$L^2$-functions $g$ and $h$:
$$
\int dx dx' g^*(x) h(x') [\Psi^*_{\ve}(x),\Psi_{\ve}(x')]_\alpha = 
\int dx dx' g^*(x) h(x') 2n(\ve)^2 \cos(\;)
\exp \left[ i \alpha j_{\vp_{\ve,x'} - \vp_{\ve,x}}\right].
$$
This converges strongly to $\langle g|h\rangle$ if for $\ve \dar 0$
$$
\left\langle\exp \left[- i \alpha j_{\vp_{\ve,x'} - \vp_{\ve,x}}\right]
\exp \left[ i \alpha j_{\vp_{\ve,y'} - \vp_{\ve,y}}\right]\right\rangle
- \left\langle\exp \left[- i \alpha j_{\vp_{\ve,x'} - \vp_{\ve,x}}\right]
\right\rangle \left\langle
\exp \left[ i \alpha j_{\vp_{\ve,y'} - \vp_{\ve,y}}\right]\right\rangle
\ra 0
$$
for almost all $x,x',y,y'$. Now
$$
\langle e^{-ij_a} e^{ij_b} \rangle = \langle e^{-ij_a}\rangle
\langle e^{ij_b}\rangle
\exp \left[\int_{-\infty}^\infty \frac{dp \;p}{1 - e^{\beta p}}
\wt a(-p) \wt b(p) \right].
$$
In our case this last factor is
\beqan
\lefteqn{ \int_{-\infty}^\infty \frac{dp \;p}{1 - e^{-\beta p}}
\frac{|1 - e^{ip\ve}|^2}{\ve^2 p^4} (e^{ipx} - e^{ipx'})
(e^{-ipy} - e^{-ipy'}) = } \\
&=& \int_{-\infty}^\infty \frac{dp\;2(1-\cos p)}{p^3(1-e^{-\beta p/\ve})}
(e^{ipx/\ve} - e^{ipx'/\ve})(e^{-ipy/\ve} - e^{-ipy'/\ve}).
\eeqan
For fixed $\beta \neq 0$ and almost all $x,x',y,y'$ this converges to zero
for $\ve \ra 0$ by Riemann-Lebesgue. In the same way one sees that
$\exp \left[i \alpha j_{\vp_{\ve,x} + \vp_{\ve,x'}} \right]$ converges
strongly to zero and that the $\Psi_{\ve,g}$ are a strong Cauchy sequence
for $\ve \ra 0$. To summarize we state
\paragraph{Theorem} (4.1) \\
$\Psi_{\ve,g}$ converges strongly for $\ve \ra 0$ to an operator
$\Psi_g$ which for $\alpha = 2\pi$ satisfies 
$$
[\Psi_g^*,\Psi_h]_+ = \langle g|h\rangle, \qquad
[\Psi_g,\Psi_h]_+ = 0.
$$
If $supp\, g < supp\, h$, 
$$
\Psi_g^* \Psi_h \, e^{i\frac{2\pi - \alpha}{4}} +
\Psi_h \Psi_g^* \, e^{-i\frac{2\pi - \alpha}{4}} = 0 \qquad \forall \alpha.
$$

Furthermore we have to verify the claim (1.5) that also for $\Psi_g$ the
current $j_f$ induces the local gauge transformation $g(x) \ra e^{2i\alpha f(x)}
g(x)$. For finite $\ve$ we have 
$$
e^{ij_f} \Psi_{\ve,g} e^{-ij_f} = 
\Psi_{\ve, e^{i 2\pi\alpha \sigma(f,\vp_\ve)}g}
$$
and for $\ve \dar 0$ we get
$\sigma(f, \vp_\ve) \ra \frac{1}{2\pi}f(0)\,$, so that $\,\sigma(f,
\tau_x\vp_\ve) = \frac{1}{2\pi}f(x)$. \\[2pt]

To conclude we investigate the status of the ``Urgleichung'' in our
construction. It is clear that the product of operator valued distributions on
the r.h.s. can assume a meaning only by a definite limiting prescription.
Formally it would be
$$
\Psi(x)\Psi^*(x)\Psi(x) = [\Psi(x), \Psi^*(x)]_+\Psi(x) - 
\Psi^*(x)\Psi(x)^2 = \delta(0)\Psi(x) - 0.
$$
From (2.11,5) we know
$$
\frac{1}{i} \frac{\partial}{\partial x} \Psi_\ve(x) = 
\frac{\sqrt{2\pi\alpha}}{2} \,[\bar \jmath(x),\Psi_\ve(x)]_+, \qquad
\bar \jmath(x) = \frac{1}{\ve} \int_{x-\ve}^x dy \; j(y).
$$
Using $j_{\vp'} e^{ij_\phi} = \frac{1}{i}\frac{\partial}{\partial\alpha}
e^{i\frac{\alpha}{2}\sigma(\vp',\vp)} e^{ij_{\vp+\alpha \vp'}}\vert_{\alpha
= 0}$ one can verify that the limit $\ve \dar 0$ exists for the expectation
value with a total set of vectors and thus gives densely defined (not
closable) quadratic forms. They do not lead to operators but we know from (2.7)
that they define operator valued distributions for test functions from $H_1$.
Thus one could say that in the sense of operator valued
distributions the Urgleichung holds
$$
\frac{1}{i} \frac{\partial}{\partial x} \Psi(x) = \frac{\sqrt{2\pi\alpha}}{2}
\,[j(x),\Psi(x)]_+ . \eqno(4.2)
$$

The remarkable point is that the coupling constant $\lambda$ in
(1.1) is related to the statistics parameter $\alpha$. For fermions one
has a solution only for $\lambda = \sqrt{2\pi}$. Of course one could for
any $\lambda$ enforce fermi statistics by renormalizing the bare
fermion field $\psi \ra \sqrt{Z}\;\psi$, $j \ra Zj$ with a suitable
$Z(\lambda)$ but this just means pushing factors around. Alternatively
one could extend $\A_c$ by adding $e^{i\sqrt{2\pi\alpha}\,j_{\vp_\ve}}$, for all
$\alpha \in {\bf R_+}$. Then one gets in $\Ha_\omega$ uncountably many
orthogonal sectors, one for each $\alpha$, and in each sector a different
Urgleichung holds. Thus different anyons live in 
orthogonal Hilbert spaces and $e^{i\sqrt{2\pi\alpha}\,j_{\vp_\ve}}$ is not even 
weakly continuous in $\alpha$. If $\alpha$ is tied to $\lambda$ it is clear that
an expansion in $\lambda$ is doomed to failure and will never reveal
the true structure of the theory.
\vspace{0.6cm}

\section{Internal symmetries}
We shall briefly discuss what happens in the the case of a fermion multiplet.
Then,
$$
\{ \psi_i^*(x), \psi_k(y)\} = \delta_{ik}\,\delta(x-y),
\qquad i = 1,2,...,N
$$
The CAR--algebra so defined posesses an obvious $U(N)$ symmetry
$$
\psi \longrightarrow U\,\psi, \quad U \in U(N)
$$
In analogy with the previous case we construct quadratic forms $\,j_k(x) =
\psi^*_k(x) \psi_k(x)\,$ which satisfy anomalous commutation relations:
\beq
[j_k(x), j_m(y)] = -\frac{i}{2\pi}\delta_{km}\delta'(x-y)
\eeq
and give rise to operators $\, j_{f_k} = \int j_k(x)f_k(x)dx, \,\, f \in H_1$.
 
Denote the corresponding Weyl operators with $W(f_1,...,f_N)$
$$
W(f_1,...,f_N) := exp\{i\sum_{n = 1}^N j_{f_n}\}
$$
The current algebra (5.1) has a (global) $O(N)$ symmetry, 
\beq
j \rightarrow  Mj = j', \qquad M \in O(N)
\eeq

Genuine anticommuting fields can now be identified 
in an extension of the current algebra quite similar to the one for the
non--chiral one--flavour case. More precisely, we have to allow for the 
existence in the new algebra of the following elements 
$$
U_{\pi_k} = e^{i2\pi\lint_{-\infty}^\infty \vp(x-x')j_k(x')dx' + 
i \dsum_{n=1\,, n \not= k}^N c_k \lint_{-\infty}^\infty j_n(x')dx'} =: \psi_k(x), 
$$
\beq
c_k \in {\bf R}, \quad  c_k - c_n = (2l+1)\pi, \quad 
\l\in{\bf Z}, \,\forall k,n 
\eeq

For the elements so defined the following relation holds
$$
\psi^{'*}_k(x) \psi'_l(y) = \psi'_l(y)\psi^{'*}_k(x)
e^{-i\pi\,\delta_{kl}\,\sgn(y-x)}\,e^{-i(1-\delta_{kl})(c_k-c_l)}
$$
which would then lead (after an appropriate renormalisation) to the desired 
CAR's. 

How should one consider the element, obtained through the same ansatz, but
after a transformation (5.2) of the currents, i.e.
\beq
\psi'_k(x) =  e^{i2\pi\int_{-\infty}^x  M_{kl}j_l(x')dx' + (...)}
\eeq
Has it something to do with the $U(N)$-transformed fermion $\psi'_k$, i.e.
$$
e^{i2\pi\int_{-\infty}^x  M_{kl}j_l(x')dx'} \, \st{?}{\longleftrightarrow} \,
U_{kl}\psi_l(x)
$$

What one notices is that the $O(N)$--transformation, being (of course) an
isomorphism of the algebra $\,\bar \A_c\,$, is no longer an automorphism of the
latter.

We are faced with the similar situation also for the $U$--invariance we
mentioned at the beginning. Here, e.g. for $\,U(1)\,$, so
$$
\psi_k(x) \rightarrow e^{i\alpha}\psi_k(x)
$$
we get 
$$
\psi'_k(x) =  e^{i\alpha}e^{i2\pi\int_{-\infty}^x j_k(x')dx'} =
e^{i2\pi\int_{-\infty}^x j'_k(x')dx'}
$$
with
$$
j'_k(x') = j_k(x') + \frac{1}{2\pi}\bar\alpha(x'), \quad
\bar\alpha_{(k)} : \int_{-\infty}^x \bar\alpha(x') = \alpha(x)
$$
In the local case this still remains an automorphism of the extended algebra,
which is in agreement with the very idea of the construction presented, while
for a global transformation this is no longer the case, so for this
fermion construction there is no global $U(1)$ symmetry present.

However, on the passage from the CAR--algebra $\,\A\,$ to the current algebra 
$\,\A_c\,$ contained in the v.Neumann algebra $\,\pi_\beta(\A)''$ the parity
has been broken, but also as an isomorphism of $\,\A_c\,$. 

Thus we are in a situation in which a particular (physically motivated)
extension $\,\bar\A_c,$ of the algebra of observables (the current algebra 
$\A_c\,$) is constructed, $\,\bar\A_c \supset \A_c$, such that 
$\,\not\exists \beta \in \mbox{Aut } \bar\A_c: \,\, \beta\vert_\A = \alpha \,$ 
for some $\, \alpha \in \mbox{Aut }\A_c\,$. This phenomenon we call 
{\it symmetry destruction}. It is related to the spontaneous collapse of a 
symmetry, discussed by Buchholz and Ojima in the context of supersymmetry 
\cite{BO} and is seen to be a field effect since in contrast to the spontaneous 
symmetry breaking \cite{NT}, it cannot occur in a finite--dimensional Hilbert 
space. 

\vspace{0.6cm}

\section{Concluding remarks}
To summarize we gave a precise meaning to eq.(1.2a,b,c) by starting with bare
fermions, $\A = {\rm CAR}(\bf R)$. The shift $\tau_t$ is an automorphism of $\A$
which has KMS--states $\omega_\beta$ and associated representations
$\pi_\beta$. In $\pi_\beta(\A)''$ one finds bosonic modes $\A_c$ with an
algebraic structure independent on $\beta$. Taking the crossed product with an
outer automorphism of $\A_c$ or equivalently augmenting $\A_c$ by an unitary
operator to $\bar\A_c$ we discover in $\bar \pi_\beta(\A_c)''$ anyonic modes
which satisfy the Urgleichung in a distributional sense. For special values of
$\lambda$ they are dressed fermions distinct from the bare ones. From the
algebraic inclusions CAR({\it bare}) $\subset \pi_\beta(\A)'' \supset\A_c \subset
\bar \A_c \subset \bar\pi_\beta(\bar\A_c)'' \supset$ CAR({\it dressed}) one 
concludes
that in our model it cannot be decided whether fermions or bosons are more
fundamental. One can construct the dressed fermions either from bare fermions
or directly from the current algebra and our original question remains open
like the one whether the egg or the hen was first.
\vspace{0.6cm}

\section*{Acknowledgements}
The authors are grateful to H. Narnhofer and R. Haag for helpful discussions on 
the subject of this paper.\\[2pt]

N.I. acknowledges the financial support from the ``Fonds zur F\"orderung der
wissenschaftlichen Forschung in \"Osterreich" under grant P11287--PHY and the
hospitality at the Institute for Theoretical Physics of University of Vienna.
\vspace{1cm}


\end{document}